\documentclass[prb,aps,twocolumn,showpacs,nofootinbib]{revtex4}
\usepackage{graphicx} 
\usepackage{dcolumn} 
\usepackage{bm} 
\usepackage{color}

\begin{document}

\title{Effective field theories and spin-wave excitations in helical magnets}

\author{Alexander I. Milstein}
\email[E-mail:]{milstein@inp.nsk.su}
\affiliation{Budker Institute of Nuclear Physics, 630090 Novosibirsk, Russia}
\author{Oleg P. Sushkov}
\email[E-mail:]{sushkov@phys.unsw.edu.au}
\affiliation{School of Physics, University of New South Wales, Sydney 2052, Australia}

\date{\today}

\begin{abstract}
We consider two classes  of helical magnets.
The first one has magnetic ordering close to antiferromagnet and 
the second one has magnetic ordering close to ferromagnet.
The first case is relevant to cuprate superconductors and the
second case is realized in FeSrO$_3$ and FeCaO$_3$.
We derive the effective field theories for these cases and
calculate corresponding excitation spectra.
We demonstrate that the ``hourglass'' spin-wave dispersion 
observed experimentally in cuprates is a fingerprint
of the ``antiferromagnetic spin spiral state''.
We also show that quantum fluctuations are important
for the ``ferromagnetic spin spiral'',
they influence qualitative features of the spin-wave dispersion. 

\end{abstract}


\maketitle
\section{introduction}
Helical spin systems have being studied for long time.
Physics of these systems is fascinating and sometimes is
very much different from that of collinear spin systems. 
The noncollinear spin systems manifest quite unusual temperature-driven 
phase transitions and critical 
properties,~\cite{Garel76,Kawamura88,Diep89,Azaria90}
for a review see Ref.~\onlinecite{Kawamura98}.
Excitation spectra (magnons) in noncollinear systems are also
very different from that in collinear magnets, see, e.g.,
Ref.~\onlinecite{Chernyshev06}.
Generically there are two ways to generate a noncollinear spin
structure, spin frustration and Dzyaloshinsky-Moriya spin orbit 
interaction. 
In the present paper we consider only the spin-frustration mechanism.
It is worth to note that the
Dzyaloshinsky-Moriya mechanism leads to interesting
quantum field theories,~\cite{Belitz06,Janoschek09}
but this is outside of the scope of the present work.

The usual field theory approach to a spin helix induced by spin
frustration  is based on the following representation
of the spin ${\vec S}$, see Ref.~\onlinecite{Kawamura98}
\begin{equation}
\label{S}
{\vec S}(t,{\bf r})={\vec a}(t,{\bf r})\cos({\bf Q}\cdot{\bf r})+
{\vec b}(t,{\bf r})\sin({\bf Q}\cdot{\bf r}) \ .
\end{equation}
Here ${\bf Q}$ is the wave vector of the spin spiral and
${\vec a}(t,{\bf r})$, ${\vec b}(t,{\bf r})$ are the dynamic fields.
The nonlinear Ginzburg-Landau action describing temperature-induced
phase transitions is written in terms of the fields 
${\vec a}$ and ${\vec b}$, see review.~\cite{Kawamura98}
Typically a frustration implies that $Q \sim 2\pi/a$, where $a$
is the lattice spacing.
Typical momenta, that appear in fluctuations of the
${\vec a}$ and ${\vec b}$ fields, are much smaller than ${\bf Q}$,
and hence ${\bf Q}$ is not influenced by the fluctuations.

Sometimes, due to accidental fine tuning of parameters, the wave vector of the 
spin spiral is small, $Q \ll 2\pi/a$, i.e. the system is close to a
collinear ferromagnet.
 This happens in FeSrO$_3$ and FeCaO$_3$ compounds.~\cite{Ulrich}
In this case typical momenta of fluctuations are comparable
with $Q$,  and hence the standard approach~\cite{Kawamura98}
based on the  Ginzburg-Landau action for ${\vec a}$ and ${\vec b}$ 
is not valid.
Another example is the case of ${\bm Q}$ close to the 
vector of antiferromagnetic ordering.
On the two-dimensional (2D) square lattice this means 
${\bf Q} \to (\pi,\pi)+{\bf Q}$. 
This is the ``antiferromagnetic spin spiral'' realised in cuprate 
superconductors.~\cite{shraiman88,Milstein08}
In this case ${\bm Q}$ is small because of the small
 doping $x$, and hence, once more, the approach based
on Eq.(\ref{S}) is not applicable.

There is one more important difference of the field theories
considered in the present paper from the case (\ref{S}).
This difference concerns a direction of ${\bm Q}$.
In the cases we consider, the vector ${\bm Q}$, in spite of being small, 
is always correlated with lattice directions. For example for a
square lattice it can be directed only along a diagonal or along
a side of the square.
This implies an essential correlation between the short range dynamics, 
$\Delta x \sim a$, and the long range dynamics, $\Delta x \sim 1/Q$.
Hence, this also implies that O(3) and SO(3) nonlinear $\sigma$-models
cannot be applied for these situations.

In the present paper we consider effective quantum field theories
for helical spin magnets close to antiferromagnets and to
ferromagnets. The structure of the paper is the following.
Section II reminds the reader the logic of derivations of the effective
action for collinear antiferromagnets and ferromagnets.
In Section III we consider the ``antiferromagnetic spin spiral'' 
and demonstrate that the ``hourglass'' spin-wave dispersion 
observed in neutron scattering from cuprates  is a fingerprint
of such state. In Section IV we consider the ``ferromagnetic spin spiral''
and show that quantum fluctuations influence qualitative features of 
the spin-wave dispersion. Section V presents our conclusions.

\section{Collinear antiferromagnets and ferromagnets}
To derive the effective action for a collinear isotropic {\it antiferromagnet}
(AF) one can use the following heuristic
arguments.\footnote{A collinear state implies that the minimum
dimensionality of the system is 2D + time.}
(i) The staggered magnetization is described by a unit vector ${\vec n}$,
${\vec n}^2=1$.
(ii) The elastic energy is proportional to $({\bm \nabla}{\vec n})^2$.
(iii) The Larmor theorem: 
If there is a state of the system without magnetic field,
then the same state in a uniform magnetic  field ${\vec B}$
must precess with 
frequency $\omega=B$.\footnote{Hereafter we set $g\mu_B=1$, $g$ is the g-factor, and $\mu_B$ is Bohr magneton.}
(iv) The static energy can depend on the magnetic field only quadratically
because the system has no net magnetization. In other words, ${\vec n}$
is not a vector, it is an axis.
The only Lagrangian  that satisfies the above requirements is the
Lagrangian of the nonlinear $\sigma$-model
\begin{eqnarray} 
\label{LLs}
{\cal L}_{AF}=\frac{\chi_{\perp}}{2}
\left({\dot{\vec n}}-[{\vec n}\times {\vec B}]\right)^2-
\frac{\rho_s}{2}\left({\bm \nabla}{\vec n}\right)^2 \ ,
 \end{eqnarray}
where the coefficients are the perpendicular magnetic susceptibility 
$\chi_{\perp}$ and the spin stiffness $\rho_s$ .

To derive the  effective action for a collinear {\it ferromagnet}
one can use similar arguments.
(i) The magnetization is described by a unit vector ${\vec n}$,
${\vec n}^2=1$.
(ii) The elastic energy is proportional to $({\bm \nabla}{\vec n})^2$.
(iii)  The Larmor theorem: 
If there is a state of the system without magnetic field,
then the same state in a uniform magnetic  field ${\vec B}$
must precess with frequency $\omega=B$.
(iv) The static energy depends on the magnetic field linearly
because ${\vec n}$ is true magnetization.
The only Lagrangian  that satisfies the above requirements reads
\begin{eqnarray} 
\label{LLf}
{\cal L}_{F}&=&iS\left(\eta^{\dag}{\dot \eta}-
{\dot \eta}^{\dag}\eta\right)-
\frac{\rho_{\xi}}{2}\left({\bm \nabla}{\vec \xi}\right)^2 +
S({\vec B}\cdot{\vec \xi})
\nonumber\\
{\vec \xi}&=&\eta^{\dag}{\vec \sigma}\eta \ . 
\end{eqnarray}
Here $S$ is spin of the ferromagnet and $\eta$ is a CP$^1$ two-component spinor.
The field ${\vec \xi}$ is a bosonic field, we just use the CP$^1$
representation for the field.
Note that the difference between antiferro and ferro is only in the
point (iv): quadratic or linear dependence of energy on an external
uniform magnetic field.

It is worth reminding that Lagrangians and fields in a quantum filed theory
always depend on normalization point. Throughout the paper we use
normalization at zero momentum (infinite distance).
This implies that in the ground state the normalization is
 $|\langle {\vec n}\rangle|=|\langle {\vec \xi}\rangle|=1$.
This also implies that spin stiffnesses and the magnetic susceptibility
in (\ref{LLs}) and (\ref{LLf}) are renormalized parameters
that include all quantum fluctuations. These are not bare parameters.

In subsequent sections
we derive and analyse effective actions that describe spin spiral states
close to antiferromagnetic and ferromagnetic cases.
This means that the wave vector of the corresponding spiral is
small compared to the inverse lattice spacing.
Specifically we have in mind cuprates for the AF case and FeCaO$_3$
for the ferro case.

\section{Spin spiral close to antiferromagnet.
Bosonic field theory inspired by cuprates}

\subsection{Effective action}
Here we consider the 2D case (2D + time), the
3D generalization is straightforward.
The effective action for the spin spiral state in the
generalized $t-J$ model has been derived in Ref.~\onlinecite{Milstein08}.
The action is written in terms of the bosonic field ${\vec n}$,
describing staggered spins of the parent Mott insulator, and the fermionic
field $\psi_{\alpha}$ describing mobile holons.
The  spinor $\psi_{\alpha}$ describes pseudospin of the holon, the pseudospin
originates from two AF sublattices. The index $\alpha=a,b$ (flavor) describes
a nodal pocket where the holon is located.
The action reads
\begin{eqnarray} 
\label{eq:LL}
{\cal L}&=&\frac{\chi_{\perp}}{2}
\left({\dot{\vec n}}-[{\vec n}\times {\vec B}]\right)^2-
\frac{\rho_s}{2}\left({\bm \nabla}{\vec n}\right)^2\\
&+&\sum_{\alpha}\left\{ \frac{i}{2}
\left[\psi^{\dag}_{\alpha}{{D}_t \psi}_{\alpha}-
{({D}_t \psi_{\alpha})}^{\dag}\psi_{\alpha}\right]\right.\nonumber\\
&+&\frac{\beta}{2}
\psi^{\dag}_{\alpha}{\bm D}^2\psi_{\alpha}  
+ \sqrt{2}g (\psi^{\dag}_{\alpha}{\vec \sigma}\psi_{\alpha})
\cdot\left[{\vec n} \times ({\bm e}_{\alpha}\cdot{\bm \nabla}){\vec n}\right] \nonumber\\
&+&\left.\frac{1}{2}({\vec B}\cdot{\vec n})
\psi^{\dag}_{\alpha}({\vec \sigma}\cdot{\vec n})\psi_{\alpha}\right\} \ .
\nonumber
\end{eqnarray}
The first line in the Lagrangian corresponds to the nonlinear 
$\sigma$-model (\ref{LLs}) describing spins.
The second line represents time derivatives of fermionic fields.
The first term in the third line is the kinetic energy of fermions,
$\beta$ is the inverse mass, the second term in the third line was derived 
long time ago by Shraiman and Siggia.~\cite{shraiman88} 
The last line represents the interaction of the pseudospin
${\vec \sigma}$ with an external uniform magnetic field.~\cite{Braz}
The unit  vectors ${\bf e}_{a}=(1/\sqrt{2}, 1/\sqrt{2})$ and
${\bf e}_{b}=(1/\sqrt{2}, -1/\sqrt{2})$ are
directed  along the corresponding diagonal of the square lattice.
The original t-J model is gauge invariant.
Therefore the effective action (\ref{eq:LL}) is also
gauge invariant and this is reflected in covariant
derivatives  
\begin{eqnarray}
\label{long}
&&{D}_t=\partial_t
+\frac{i}{2}{\vec \sigma}\cdot[{\vec n}\times{\dot{\vec n}}] \nonumber\\
&&{\bm D}={\bm \nabla}
+\frac{i}{2}{\vec \sigma}\cdot[{\vec n}\times{\bm \nabla}{\vec n}] \ ,
\end{eqnarray}
acting on the fermionic field.

The Lagrangian (\ref{eq:LL}) describes many physical effects:
superconductivity, spin spiral, particle-hole decay, even phase separation
for some values of parameters.
In the present work we aim only at the spin spiral.
Therefore we will truncate fermionic degrees of freedom 
in (\ref{eq:LL}) throwing away all physical effects except of the spin spiral.
We retain only spin  and pseudospin degrees of freedom responsible for the
spin spiral.
The way of truncation is pretty straightforward, we replace the
fermionic field $\psi_{\alpha}$ by a bosonic field $\xi_{\alpha}$ describing \
pseudospin in the  CP$^1$ representation 
\begin{equation}
\label{trunc}
\psi^{\dag}_{\alpha}{\vec \sigma}\psi_{\alpha} \to{\vec \xi}_{\alpha}
=\eta^{\dag}_{\alpha}{\vec \sigma}\eta_{\alpha} \ .
\end{equation}
Hence, the truncated Lagrangian reads
\begin{eqnarray} 
\label{AFs}
{\cal L}_{AFS}&=&\frac{\chi_{\perp}}{2}
\left({\dot{\vec n}}-[{\vec n}\times {\vec B}]\right)^2-
\frac{\rho_s}{2}\left({\bm \nabla}{\vec n}\right)^2\nonumber\\ 
&+&S_{\xi}\sum_{\alpha}\left\{\frac{i}{2}
\left(\eta^{\dag}_{\alpha}{D_t \eta}_{\alpha}-{( D_t \eta_{\alpha})}^{\dag}
\eta_{\alpha}\right)\right.\nonumber\\
&-&\frac{\rho_{\xi}}{2}\left({\hat {\bm D}}{\vec \xi}_{\alpha}\right)^2
+\frac{g}{\sqrt{2}}
\left({\vec \xi}_{\alpha}\cdot[{\vec n}\times({\bm e}_{\alpha}\cdot{\bm \nabla})
{\vec n}]\right)\nonumber\\
&+&\left.\frac{1}{2}({\vec B}\cdot{\vec n})({\vec \xi}_{\alpha}\cdot{\vec n})
\right\} \ ,
\end{eqnarray}
where $S_{\xi}$ is the effective ``spin'' of the $\xi$-field,
the stiffness of the $\xi$-field is $\rho_{\xi}$.
The coupling constant between ${\vec n}$ and ${\vec \xi}$ fields is $g$,
it does not coincide with $g$ in (\ref{eq:LL}), but they are related.
Importantly, the truncated Lagrangian (\ref{AFs}) still respects the gauge
invariance of the original t-J model since it contains only covariant
derivatives acting on the pseudospin, ${D}_t$ is defined in Eq.(\ref{long}),
and the spacial derivative is defined as
\begin{eqnarray}
\label{long1}
{\hat {\bm D}}{\vec \xi}_{\alpha}={\bm \nabla}{\vec \xi}_{\alpha}
+{\vec \xi}_{\alpha}\times[{\vec n}\times{\bm \nabla}{\vec n}] \ .
\end{eqnarray}
A related important point is that (\ref{AFs}) respects also the Larmor
theorem. If there is a state without magnetic field that satisfies
equations of motion, then in a magnetic field this state precesses 
with frequency $\omega=B$. This is a sort of a global Ward identity.

It is worth noting that statuses of the Lagrangians (\ref{eq:LL}) and
(\ref{AFs}) are different. The Lagrangian (\ref{eq:LL})
has been derived from the generalized t-J model in a controlled
way~\cite{Milstein08}. The small parameter that controls
accuracy of the derivation is small doping.
On the other hand we obtain the Lagrangian (\ref{AFs}) from
(\ref{eq:LL}) by an uncontrolled truncation of fermionic
degrees of freedom. As a result we get an effective Ginzburg-Landau
action (\ref{AFs}) that contains only spin ${\vec n}$
and pseudospin ${\vec \xi}$.
In the present paper we analyze only the action (\ref{AFs})
which, as we show below, describes experimental observations
remarkably well.

The conventional approach~\cite{Kawamura98} to the spin spiral state 
is based on Eq.~(\ref{S}),  it contains two vector dynamic fields 
${\vec a}(t,{\bf r})$ and  ${\vec b}(t,{\bf r})$.
Our approach  also contains two vector dynamic fields
${\vec n}(t,{\bf r})$ and ${\vec \xi}(t,{\bf r})$.
The difference is that our effective action (\ref{AFs})
is valid for any $q \ll 1$, it is not restricted by the region 
$q \ll Q$. We will show that in the limit $q \ll Q$ our approach 
results in two independent Goldstone modes with linear dispersions.
Hence, as one should expect, in this limit both approaches are equivalent.

Hereafter we set the external magnetic field equal to zero, $B=0$.
Neglecting quantum fluctuations one can easily find the ground state
of the Lagrangian (\ref{AFs}),
\begin{eqnarray}
\label{Qn}
{\vec n}&=&(C,S,0)\nonumber\\
C&=&\cos({\bm Q}\cdot{\bm r})\nonumber\\
S&=&\sin({\bm Q}\cdot{\bm r})\nonumber\\
{\vec \xi}_a&=&{\vec \xi}_b=(0,0,1)\nonumber\\
{\bm Q}&=&\frac{g}{\rho_s}\frac{{\bm e}_a+{\bm e}_b}{\sqrt{2}}
=\frac{g}{\rho_s}(1,0)  \ .
\end{eqnarray}
Here we choose x- and y-directions in the spin space to lie in the
plane of the spin spiral.
Both ${\vec \xi}_a$ and ${\vec \xi}_b$ are directed along z in the spin space, 
and in this case the wave vector of the spiral ${\bm Q}$ is directed along 
the (1,0) direction of the square lattice.
One can also choose ${\vec \xi}_a=-{\vec \xi}_b=(0,0,1)$.
In this case ${\bm Q}\propto {\bm e}_a-{\bm e}_b$ 
is directed along the $(0,1)$ direction of the square 
lattice. The choice between these two possibilities is spontaneous.

\subsection{The in-plane spin-wave excitation}
Consider the spin-wave excitations above the ground state (\ref{Qn}).
There is the in-plane excitation $\varphi=\varphi(t,{\bf r})$, where
\begin{eqnarray}
\label{inp}
&&{\vec n}=(\cos({\bf Q\cdot r}+\varphi),\sin({\bf Q\cdot r}+\varphi),0) 
\nonumber\\
&&{\vec \xi}_{\alpha}=(0,0,1) \ ,
\end{eqnarray}
and the out-of-plane excitation $n_z$, ${\vec \xi}_{\perp}$, where
\begin{eqnarray}
\label{Qnz}
{\vec n}&=&(C,S,0)\sqrt{1-n_z^2}
+(0,0,n_z)\nonumber\\
{\vec \xi}_{\alpha}&=&(\xi_{\alpha x},\xi_{\alpha y},\sqrt{1-\xi_{\alpha\perp}^2})
\nonumber\\
{\vec \xi_{\alpha\perp}}&=&(\xi_{\alpha x},\xi_{\alpha y},0) \ .
\end{eqnarray}
Note that we use indexes $x,y,z$ to denote axes in the
spin space and we use indexes 1,2 to denote directions
in the coordinate space, the direction 1 is parallel to the spin spiral
vector ${\bm Q}$ and the direction 2 is perpendicular to the spin spiral
vector.

We first consider the in-plane magnon (\ref{inp}) that does not involve
fluctuations of the $\xi$-field.
Substituting expression (\ref{inp}) in the Lagrangian (\ref{AFs})
and keeping only quadratic terms in the small perturbation $\varphi$
one finds
\begin{eqnarray}
\label{LNf}
{\cal L}\to
\frac{\chi_{\perp}}{2}{\dot{\varphi}}^2-
\frac{\rho_s}{2}\left({\bm \nabla}{\varphi}\right)^2 \ .
\end{eqnarray}
Hence the Green's function of the in-plane magnon 
is the same as the  Green's function of a magnon in the parent
$\sigma$-model,
\begin{equation} 
\label{Gin}
G_{\varphi}(\omega,{\bm q})=\frac{\chi_{\perp}^{-1}}{\omega^2-c^2q^2+i0} \ ,
\end{equation}
where
\begin{equation}
\label{cc}
c=\sqrt{\frac{\rho_s}{\chi_{\perp}}} 
\end{equation}
is the bare spin-wave speed.
\begin{figure}[h!]
\includegraphics[width=0.25\textwidth,clip]{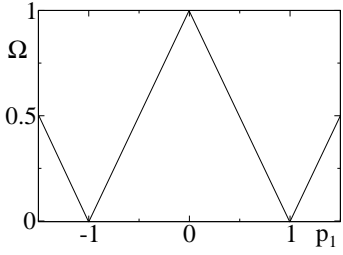}
\caption{The spin wave dispersion of the in-plane magnon as it is
seen in neutron scattering
}
\label{IntFI}
\end{figure}
The in-plane magnons can be excited in inelastic neutron scattering.
Neutrons interact only with the $n$-field that represents real
spins, therefore the scattering
amplitude is given by the Fourier transform of ${\vec n}$ from (\ref{inp}) and the
scattering probability for unpolarized neutrons with the
energy transfer $\omega$ and the momentum transfer ${\bm k}$
is proportional to
\begin{eqnarray} 
\label{i1}
{I}_{in}(\omega,{\bf k})=-\frac{1}{2\pi}{\text {Im}}
\left[ G_{\varphi}(\omega,{\bf k}-{\bf Q})
+ G_{\varphi}(\omega,{\bf k}+{\bf Q})\right]  
\end{eqnarray}
Hereafter we introduce the dimensionless energy and 
the dimensionless momentum
\begin{eqnarray}
\label{dim}
\Omega=\frac{\omega}{cQ} \ , \ \ \ \ \ {\bm p}=\frac{{\bm q}}{Q} \ .
\end{eqnarray}
The momentum is measured in units of the wave vector of the
spin spiral and the energy is measured in units of the magnon energy
in the parent $\sigma$-model with momentum of the magnon equal to
the spin spiral wave vector.
According to Eq.(\ref{i1}) the in-plane magnon is seen in neutron scattering
as two cones originating from the points $\pm {\bm Q}$.
The corresponding plot for the direction along the spin spiral
is shown in Fig.~\ref{IntFI}.

\subsection{The out-of-plane spin-wave excitation}

We have shown in the previous subsection that
the in-plane magnon in the spin-spiral state remains trivial.
It is not the case for the out-of-plane magnon.
Substitution of (\ref{Qnz}) in Eq.(\ref{AFs}) gives the
following action expressed in terms of $n_z$ and ${\vec \xi}_{\alpha\perp}$
\begin{eqnarray} 
\label{AFs1}
&&{\cal L}_{AFS}=\frac{\chi_{\perp}}{2}{\dot n}_z^2-
\frac{\rho_s}{2}[Q^2(1-n_z^2)+({\bm \nabla}n_z)^2] \\
&+&S_{\xi}\sum_{\alpha}\left\{\frac{i}{2}
\left(\eta^{\dag}_{\alpha}{\dot \eta}_{\alpha}-{\dot \eta}^{\dag}_{\alpha}
\eta_{\alpha}\right)-\frac{1}{2}(S\xi_{\alpha x}-C\xi_{\alpha y})
{\dot n}_z\right\}
\nonumber\\
&-&\frac{\rho_{\xi}}{2}\sum_{\alpha}\left[
\left({\bm \nabla}\xi_{\alpha x}+{\bm Q}\xi_{\alpha y}
+S{\bm Q}n_z+C{\bm \nabla}n_z\right)^2\right.\nonumber\\
&+&\left.\left({\bm \nabla}\xi_{\alpha y}-{\bm Q}\xi_{\alpha x}
-C{\bm Q}n_z+S{\bm \nabla}n_z\right)^2\right]
\nonumber\\
&+&\frac{g}{\sqrt{2}}\sum_{\alpha}{\bm e}_{\alpha}
\left[{\bm Q}\left(1-n_z^2-\frac{1}{2}\xi_{\alpha\perp}^2\right)
\right.\nonumber\\
&+&\left.\left(-C\xi_{\alpha x}-S\xi_{\alpha y}\right){\bm Q}n_z
+\left(S\xi_{\alpha x}-C\xi_{\alpha y}\right){\bm \nabla}n_z\right]\ .
\nonumber
\end{eqnarray}
We keep here only quadratic terms.
Let us perform in (\ref{AFs1}) two subsequent gauge transformations.
The first one is
\begin{eqnarray}
\label{gt1}
\eta_{\alpha}&=&\exp\left\{-\frac{i}{2}({\bm Q}\cdot{\bm r})\sigma_z\right\}
\eta'_{\alpha}
\nonumber\\
\xi_{\alpha x}&=&C\xi_{\alpha x}'-S\xi_{\alpha y}'\nonumber\\
\xi_{\alpha y}&=&S\xi_{\alpha x}'+C\xi_{\alpha y}' \ .
\end{eqnarray}
The second  transformation reads
\begin{eqnarray}
\label{gt2}
\eta'_{\alpha}&=&\exp\left\{\frac{i}{2}n_z\sigma_y\right\}
\eta_{\alpha}''\nonumber\\
\xi_{\alpha x}'&\approx&\xi_{\alpha x}''-n_z\nonumber\\
\xi_{\alpha y}'&=&\xi_{\alpha y}''\ .
\end{eqnarray}
Geometrically the gauge transformations (\ref{gt1}) and (\ref{gt2})
describe the transition to the proper reference frame of the vibrating
spin spiral.
In terms of the new variables $\eta''$ and $\xi''$ the Lagrangian (\ref{AFs1})
 consists of three parts
\begin{eqnarray}
\label{s3}
&&{\cal L}_{AFS}={\cal L}_{n}+{\cal L}_{\xi}+{\cal L}_{int}\\
&&{\cal L}_{n}=\frac{\chi_{\perp}}{2}{\dot n}_z^2-
\frac{\rho_s}{2}({\bm \nabla}n_z)^2-\rho_{\xi} Q^2n_z^2\nonumber\\
&&{\cal L}_{\xi}=\sum_{\alpha}\left\{S_{\xi}\frac{i}{2}
\left(\eta^{\dag}_{\alpha}{\dot \eta}_{\alpha}-{\dot \eta}^{\dag}_{\alpha}
\eta_{\alpha}\right)
\right.\nonumber\\
&&\ \ \ \ -\left.\frac{\rho_{\xi}}{2}\left[({\bm \nabla}\xi_{\alpha x})^2+
({\bm \nabla}\xi_{\alpha y})^2
\right] -\rho_s\frac{Q^2}{4}\xi_{\alpha\perp}^2\right\} \nonumber\\
&&{\cal L}_{int}=
\sum_{\alpha}\left\{-\frac{g}{\sqrt{2}}\xi_{\alpha y}
({\bm e}_{\alpha }\cdot{\bm \nabla})n_z
+ \rho_{\xi} n_z({\bm Q}\cdot{\bm \nabla})\xi_{\alpha y}
\right\} \ . \nonumber
\end{eqnarray}
Hereafter we skip the double prime in notations of $\eta''$ and $\xi''$.
The first part, ${\cal L}_{n}$, describes the free motion of the $n_z$-field;
the second part, ${\cal L}_{\xi}$, describes the free motion of the $\xi$-fields;
and the third part, ${\cal L}_{int}$, describes the interaction between $n_z$ and $\xi$.
Equations of motion generated by the action (\ref{s3}) are
\begin{eqnarray}
\chi_{\perp}{\ddot n}_x&=&\rho_s\Delta n_z -2\rho_{\xi}Q^2n_z\nonumber\\
&+&\sum_{\alpha}\left[\frac{g}{\sqrt{2}}
({\bm e}_{\alpha}\cdot{\bm \nabla})\xi_{\alpha y}+
\rho_{\xi}({\bm Q}\cdot{\bm\nabla})\xi_{\alpha y}\right]\nonumber\\
S_{\xi}{\dot \xi}_{\alpha x}&=&-2\left[
\rho_{\xi}\Delta\xi_{\alpha y}-\frac{\rho_sQ^2}{2}\xi_{\alpha y}
\right.\nonumber\\
&-&\left. \frac{g}{\sqrt{2}}({\bm e}_{\alpha}\cdot{\bm \nabla})n_z
-\rho_{\xi}({\bm Q}\cdot{\bm\nabla})n_z\right]\nonumber\\
S_{\xi}{\dot \xi}_{\alpha y}&=&2\left[
\rho_{\xi}\Delta\xi_{\alpha x}-\frac{\rho_sQ^2}{2}\xi_{\alpha x}
\right] \ .
\end{eqnarray}
Looking for solution in the form
\begin{eqnarray}
\label{sol1}
&&n_z=Ae^{-i\omega t+i{\bm q}\cdot{\bm r}}\nonumber\\
&&\xi_{\alpha x}=B_{\alpha}e^{-i\omega t+i{\bm q}\cdot{\bm r}}\nonumber\\
&&\xi_{\alpha y}=-iC_{\alpha}e^{-i\omega t+i{\bm q}\cdot{\bm r}}
\end{eqnarray}
we find the relations between the coefficients $A$, $B_{\alpha}$, $C_{\alpha}$,
and the frequencies of excitations,
\begin{eqnarray}
\label{om}
&&\Omega_{1,\bm p}^2,\Omega_{2,\bm p}^2
=\frac{1}{2}(O_1+O_2)\mp\sqrt{\frac{1}{4}(O_1-O_2)^2+O_3}  
\nonumber \\
&&O_1 =p^2+b^2 \nonumber\\
&&O_2=a^2\left(1+b^2p^2\right)^2 \nonumber\\
&&O_3=a^2\left(1+b^2p^2\right)\left[p_1^2(1+b^2)^2+p_2^2\right]  \nonumber\\
&&p^2=p_1^2+p_2^2 \ .
\end{eqnarray}
These equations are written in terms of dimensionless
frequency and momentum (\ref{dim}). We remind that $p_1$
is parallel to ${\bm Q}$ and $p_2$ is perpendicular to ${\bm Q}$.
In (\ref{om}) we have introduced two dimensionless parameters
\begin{eqnarray}
\label{ab}
a=\frac{Q\rho_s}{cS_{\xi}}\ , \ \ \ \ b=\sqrt{\frac{2\rho_{\xi}}{\rho_s}} \ .
\end{eqnarray}
Note that these parameters are unrelated to the a,b-flavors.

\begin{figure}[h]
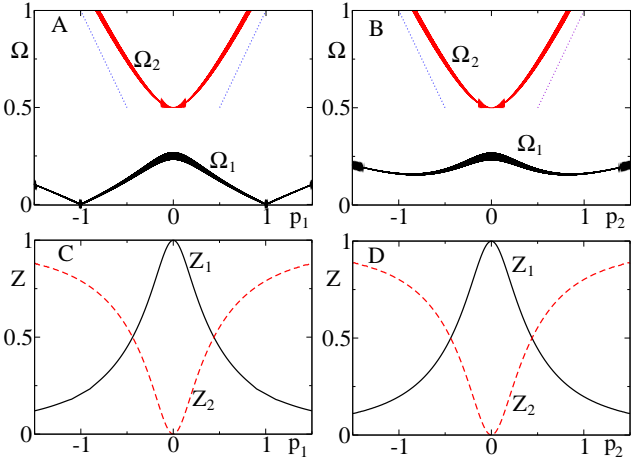

\includegraphics[width=0.23\textwidth,clip]{dispA05.1.eps}
\includegraphics[width=0.23\textwidth,clip]{dispA05.2.eps}
\includegraphics[width=0.23\textwidth,clip]{ZA05.1.eps}
\includegraphics[width=0.23\textwidth,clip]{ZA05.2.eps}
\caption{(Color online) A and B: the dispersion $\Omega_{\bm k}$
of the out-of-plane magnon 
in directions parallel to the spin spiral, (A);  and  perpendicular
to the spin spiral, (B). Values of parameters of the effective field theory
are $a=0.5$, $b=0.25$.
The blue dotted lines show the magnon dispersion in the
parent $\sigma$-model.
Thickness of $\Omega_1$ and $\Omega_2$ lines is proportional to the
corresponding quasiparticle residue $Z_{1}$, $Z_{2}$.
The quasiparticle residues  are plotted separately
for the same  directions in Figs. C  and D. 
}
\label{dispA05}
\end{figure}

\begin{figure}[h]
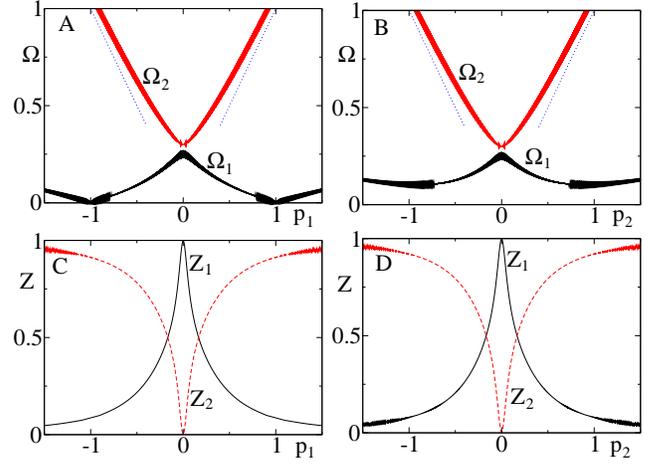

\includegraphics[width=0.23\textwidth,clip]{dispA03.1.eps}
\includegraphics[width=0.23\textwidth,clip]{dispA03.2.eps}
\includegraphics[width=0.23\textwidth,clip]{ZA03.1.eps}
\includegraphics[width=0.23\textwidth,clip]{ZA03.2.eps}
\caption{(Color online) A and B: the dispersion $\Omega_{\bm k}$
of the out-of-plane magnon 
in directions (A) parallel to the spin spiral,  and (B) perpendicular
to the spin spiral. Values of parameters of the effective field theory
are $a=0.3$, $b=0.25$.
The blue dotted lines show the magnon dispersion in the
parent $\sigma$-model.
Thickness of $\Omega_1$ and $\Omega_2$ lines is proportional to the
corresponding quasiparticle residue $Z_{1}$, $Z_{2}$.
The quasiparticle residues  are plotted separately
for the same  directions in Figs. C  and D. 
}
\label{dispA03}
\end{figure}

Importantly, there are two branches of the dispersion that
originate from two different degrees of freedom, spin ${\vec n}$ and 
pseudospin ${\vec \xi}$.
In order to reproduce the spin-wave excitation spectra
 of the t-J model at small doping
we chose $b << 1$, $a > b$.
The parameter $b$ defined in (\ref{ab}) determines the top of
the lower branch $\Omega_{1,\bm p}$, and
the parameter $a$ defined in (\ref{ab}) determines the bottom of
the upper branch $\Omega_{2,\bm p}$.
Plots of $\Omega_{1,\bm p}$ and $\Omega_{2,\bm p}$, for directions
of ${\bm p}$ parallel to the spin spiral and perpendicular to the
spin spiral  are presented  in Figs.~\ref{dispA05}A,B
for values of parameters $b=0.25$,  $a=0.5$.
Similar plots for $b=0.25$,  $a=0.3$ are presented 
in Figs.~\ref{dispA03}A,B.
In Figs. \ref{dispA05} and \ref{dispA03} the dispersion is shown
by very thick lines. The thickness is not a decay width, the decay width
in this approximation (no alive fermions) is negligible.
The thickness is proportional to the corresponding quasiparticle residue
calculated below.

To find absolute values of the coefficients  $A$, $B_{\alpha}$, $C_{\alpha}$,
in (\ref{sol1}) we need to quantize (\ref{s3}). 
There are two magnon creation operators, $c_{1,{\bm p}}^{\dag}$, and
$c_{2,{\bm p}}^{\dag}$, corresponding to two branches of the dispersion.
The standard quantization gives
\begin{eqnarray}
\label{sol2}
&&n_z=\sum_{l=1,2} N_{l,{\bm p}} +h. c.\\
&&N_{l,{\bm p}}=\frac{1}{\sqrt{\chi_{\perp}cQ}}
\sqrt{\frac{Z_{l,{\bm p}}}{2\Omega_{l,{\bm p}}}} \ 
c_{l {\bm p}}e^{-i\omega t+i{\bm q}\cdot{\bm r}}\nonumber\\
&&\xi_{\alpha x}=-\sum_{l=1,2}
\frac{a\Omega_{l,{\bm p}}
\left[\sqrt{2}({\bm e}_{\alpha}\cdot{\bm p})+b^2p_1\right]}
{\Omega_{l,{\bm p}}^2-O_2} \ N_{l,{\bm p}} + h.c.\nonumber\\
&&\xi_{\alpha y}=i\sum_{l=1,2}
\frac{a^2(1+b^2p^2)\left[\sqrt{2}({\bm e}_{\alpha}\cdot{\bm p})+b^2p_1\right]}
{\Omega_{l,{\bm p}}^2-O_2} \ N_{l,{\bm p}} + h.c.\nonumber
\end{eqnarray}
Here index $l=1,2$ enumerates the dispersion brunches,
the expression for $O_2$ is given in (\ref{om}), and $Z_{l,{\bm p}}$ is
the quasiparticle residue
\begin{eqnarray}
\label{Z12}
&&Z_{1,{\bm p}},Z_{2,{\bm p}}=\frac{1}{2}\left\{1\mp\frac{(O_1-O_2)}{
\sqrt{(O_1-O_2)^2+4 O_3}}\right\}\ .
\end{eqnarray}
The quasiparticle residues $Z_{1,{\bf p}}$ and $Z_{2,{\bf p}}$ are plotted
in Figs.~\ref{dispA05} C,D and in Figs.~\ref{dispA03} C,D.
Interestingly, the quasiparticle residues are practically
isotropic, the dependence on the direction of ${\bf p}$ is
extremely weak.
The magnon Green's function defined in the standard way reads
\begin{eqnarray}
\label{nq1}
G_{n_z}(\omega,{\bm q})&=&-i\int\langle T
\left\{n_z({\bm r},t)n_z(0,0)\right\}\rangle
e^{i\omega t-i{\bm q}\cdot{\bm r}}dtd^2r\nonumber\\
&=&
\frac{\chi_{\perp}^{-1}Z_{1,{\bm q}}}{\omega^2-\omega_{1,\bm q}^2+i0} 
+
\frac{\chi_{\perp}^{-1}Z_{2,{\bm q}}}{\omega^2-\omega_{2,\bm q}^2+i0} \ . 
\end{eqnarray}
We remind that $\omega$, $\Omega$ and  ${\bm q}$, ${\bm p}$ are
related according to Eq.(\ref{dim}). 
The inelastic neutron scattering intensity for unpolarized 
neutrons with the
energy transfer $\omega$ and the momentum transfer ${\bm k}$
is proportional to
\begin{eqnarray} 
\label{oi1}
{I}_{out}(\omega,{\bf k})=-\frac{1}{\pi}{\text {Im}}\ G_{n_z}(\omega,{\bf k})
\end{eqnarray}
 
Figs. \ref{dispA05} and \ref{dispA03} show the ``hourglass''
dispersion that agrees with the typical  dispersion observed in 
cuprates.~\cite{Tranq,hinkov04,hinkov07a} 
The lower branch $\Omega_{1,{\bm q}}$ has strongly anisotropic
momentum dependence with zeros at
the Goldstone points ${\bm q}=\pm{\bm Q}$, while the
upper branch $\Omega_{2,{\bm q}}$ is practically isotropic in
the momentum space quickly approaching the bare spin-wave dispersion 
at larger $q$.
The intensity (quasiparticle residue) of the
lower branch $Z_{1,{\bm q}}$ is strongly peaked at ${\bm q}=0$, at
the ``neck'' of the ``hourglass'',  and the intensity decays 
dramatically away from this point, see Figs.~\ref{dispA05} C,D and  
Figs.~\ref{dispA03} C,D.
On the other hand the intensity of the
upper branch $Z_{2,{\bm q}}$ is zero exactly at ${\bm q}=0$
and quickly approaches unity away from this point.
The  dispersion obtained in the present work demonstrates that
the ``hourglass'' shape is a ``fingerprint'' of the spin spiral
driven by the interaction between spin and pseudospin. The two 
branches $\Omega_1$
and $\Omega_2$ reflect the two degrees of freedom: spin and pseudospin.

\subsection{Quantum fluctuations}
The spin spiral state that we consider can be characterized by
the  vector order parameters,
that indicate ordering of spin ${\vec n}$ and  pseudospin 
${\vec \xi}$, and
a pseudoscalar order parameter ${\cal P}$ that indicates chirality.
The order parameters are
\begin{eqnarray}
\label{o1}
&&\langle{\vec n}\rangle \ , \ \ \ \langle{\vec \xi}_a\rangle
\ , \ \ \ \langle{\vec \xi}_b\rangle \nonumber \\
&&{\cal P}=\langle\frac{1}{\sqrt{2}}\sum_{\alpha}{\vec \xi}_{\alpha}\cdot
[{\vec n}\times({\bm e}_{\alpha}\cdot{\bm \nabla}){\vec n}]\rangle \ .
\end{eqnarray}
In semiclassical approximation values of the vector order parameters are
given by Eqs. (\ref{Qn}), the chiral pseudoscalar is ${\cal P}=Q$.
Note that ${\cal P}$ changes sign under time reflection.

Due to quantum fluctuations values of some order parameters are
reduced. The reduction of the spin field is
\begin{equation}
\label{nred}
|\langle {\vec n} \rangle|=\langle\sqrt{1-\varphi^2-n_z^2}\ \rangle
 \approx 1-\frac{1}{2}\langle \varphi^2\rangle
-\frac{1}{2}\langle n_z^2\rangle \ .
\end{equation}
Only the length is reduced, the spacial dependence is still
given by Eq. (\ref{Qn}), 
$\langle {\vec n} \rangle= |\langle n \rangle|
(\cos({\bm Q}\cdot{\bm r}),\sin({\bm Q}\cdot{\bm r}), 0)$.
To calculate the reduction of the length we note that
the expectation values $\langle \varphi^2\rangle$ and
$\langle n_z^2\rangle$ can be expressed in terms of the Green's function,
\begin{eqnarray}
\label{fin}
\langle \varphi^2\rangle&=&-\sum_{\bf q}\int\frac{d\omega}{2\pi i}
G_{in}(\omega,{\bf q})\nonumber \\
\langle n_z^2\rangle&=&-\sum_{\bf q}\int\frac{d\omega}{2\pi i}
G_{out}(\omega,{\bf q}) \ .
\end{eqnarray}
The physical meaning of relations (\ref{nred}) and (\ref{fin}) is very simple: 
the reduction of static response is transferred to the dynamic response.
The expressions (\ref{nred}) and (\ref{fin})
must be renormalized by subtraction of the 
ultraviolet-divergent contribution that corresponds to the parent 
$\sigma$-model. Since the Green's function of the in-plane excitation
(\ref{Gin}) is exactly the same as the magnon Green's function in the
parent  $\sigma$-model, the fluctuation $\langle \varphi^2\rangle$
is renormalized to zero. The out-of plane fluctuation gives
\begin{eqnarray}
\label{nred1}
|\langle {\vec n} \rangle| &\approx& 1-\frac{1}{2}\langle n_z^2\rangle=
1-\frac{cQ}{2\rho_s}{\cal I}_R\nonumber\\
{\cal I}_R&=&\int\left\{\frac{Z_{1,{\bm p}}}{2\Omega_{1,{\bm p}}}
+\frac{Z_{2,{\bm p}}}{2\Omega_{2,{\bm p}}}-\frac{1}{2p}\right\}\frac{d^2p}{(2\pi)^2}
 \ .
\end{eqnarray}
Here the subscript $R$ stands for ``renormalized'', ${\cal I} \to {\cal I}_R$ .
The dimensionless integral ${\cal I}_R$ can be easily calculated, it depends
only on the parameters $b$ and $a$ defined in (\ref{ab}).
Values of the integral for $a$ and $b$ corresponding to 
Figs.~\ref{dispA05},\ref{dispA03} are 
\begin{eqnarray}
\label{IR}
&&{\cal I}_R(a=0.5,b=0.25)=0.23 \nonumber\\ 
&&{\cal I}_R(a=0.3,b=0.25)=0.15 \ .
\end{eqnarray}

The pseudospin ${\vec \xi}_{\alpha}$ is not a gauge invariant object,
it cannot be directly measured and its expectation value depends on gauge.
One can calculate the expectation value of $ {\vec \xi}_{\alpha} $
defined in the original action  (\ref{AFs}) or one can calculate
the expectation value of ${\vec \xi}_{\alpha}'' $ obtained after the
gauge transformation (\ref{gt1}), (\ref{gt2}).
Below we calculate $\langle (\xi_{\alpha\perp}'')^2\rangle $, but we skip the double 
prime in notations.
Using Eqs. (\ref{sol2}) we find
\begin{eqnarray}
\label{xired1}
&&\langle \xi_{\alpha \perp}^2 \rangle =
\langle \xi_{\alpha x}^2 \rangle +\langle \xi_{\alpha y}^2 \rangle 
=\frac{cQ}{\rho_s}{\cal J}\\
&&\hspace{-20pt}{\cal J}=\int\sum_{l=1,2}\frac{Z_{l,{\bm p}}}{2\Omega_{l,{\bm p}}}
\frac{a^2[\Omega_{l,{\bm p}}^2+O_2][p_1^2(1+b^2)^2+p_2^2]}
{[\Omega_{l,{\bm p}}^2-O_2]^2}\frac{d^2p}{(2\pi)^2} \ . \nonumber
\end{eqnarray}
At $p \to\infty$ the integrand in ${\cal J}$
is approaching the constant value equal to $a$. 
This corresponds to the ultraviolet divergence of 
$\langle \xi_{\alpha }^2 \rangle$ in the parent $\xi$-model
defined by Eq. (\ref{LLf}).
Therefore ${\cal J}$ must be renormalized, 
${\cal J}\to {\cal J}_R$, in the following way
\begin{eqnarray}
\label{xiJ}
\hspace{-20pt}{\cal J}_R=\int\left\{\left(...\right)
-a\right\}
\frac{d^2p}{(2\pi)^2} \ ,
\end{eqnarray}
where $\left(...\right)$ stands for the integrand in (\ref{xired1}).
After the renormalization ${\cal J}_R$ is convergent.

The chiral pseudoscalar order parameter ${\cal P}$ is defined in Eq. (\ref{o1}).
In this definition the field $\xi_{\alpha}$ is the pseudospin of the 
original action (\ref{AFs}),
and in this sense the chiral order parameter is gauge invariant.
It is easy to rewrite the order parameter in terms of the gauge transformed
$\xi_{\alpha}''$ and $n_z$, 
\begin{eqnarray}
\label{o2}
\frac{{\cal P}}{Q}&=&
1-\frac{1}{2}\langle n_z^2\rangle
-\frac{1}{2}\langle \xi_{\alpha \perp}^2\rangle
-\frac{1}{\sqrt{2}Q}
\langle\sum_{\alpha}\xi_{\alpha y}({\bm e}_{\alpha}\cdot{\bm \nabla})n_z\rangle \ ,
\nonumber\\
\end{eqnarray}
where we again omit the double prime in notations.
The first two terms after the unity have been already
calculated in Eqs.(\ref{nred1}),(\ref{xired1}), and the last term
immediately follows from (\ref{sol2}),
\begin{eqnarray}
\label{xinr}
&&\frac{1}{\sqrt{2}Q}\langle \sum_{\alpha}{\bm e}_{\alpha}
\xi_{\alpha y}{\bm \nabla}n_z\rangle = \frac{cQ}{2\rho_s}{\cal K}\\
&&{\cal K}=\int\sum_{l=1,2}\frac{Z_{l,{\bm p}}}{2\Omega_{l,{\bm p}}}
\frac{2a^2[1+b^2p^2][p^2+b^2p_1^2]}
{[\Omega_{l,{\bm p}}^2-O_2]}\frac{d^2p}{(2\pi)^2} \ . \nonumber
\end{eqnarray}
The integrand in ${\cal K}$ decays at large $p$ as $\propto 1/p$,
so ${\cal K}$ is ultraviolet divergent. We will discuss this
divergence in subsection E.

Quantum fluctuations do not influence direction of the
spin spiral wave vector ${\bm Q}$, but they change the absolute value
of the vector.
To calculate the quantum fluctuation correction to Q
 we need to account for nonlinear terms in the
Lagrangian (\ref{AFs}).
To do so it is convenient to generalize the notations (\ref{inp}) and 
(\ref{Qnz}) for the in-plane and out-of-plane excitations
in the following way
\begin{eqnarray}
\label{Qnzb}
{\vec n}&=&(C_{\varphi},S_{\varphi},0)C_{\gamma}
+(0,0,1)S_{\gamma} \nonumber\\
{\vec \xi}_{\alpha}&=&(\xi_{\alpha x},\xi_{\alpha y},
\sqrt{1-\xi_{\alpha \perp}^2})\nonumber\\
C_{\varphi}&=&\cos({\bm Q}\cdot{\bm r}+\varphi) \nonumber\\
S_{\varphi}&=&\sin({\bm Q}\cdot{\bm r}+\varphi) \nonumber\\
C_{\gamma}&=&\cos\gamma\nonumber\\
S_{\gamma}&=&n_z=\sin\gamma \ .
\end{eqnarray}
The gauge transformation (\ref{gt1}),(\ref{gt2}) is generalized as
\begin{eqnarray}
\label{gt4}
\eta_{\alpha}&=&
\exp\left\{-\frac{i}{2}({\bm Q}\cdot{\bm r}+\varphi)\sigma_z\right\}
\exp\left\{\frac{i}{2}\gamma\sigma_y\right\}\eta_{\alpha}''\nonumber\\
\xi_{\alpha x}&=&C_{\varphi}(C_{\gamma}\xi''_{\alpha x}-S_{\gamma}\xi''_{\alpha z})
-S_{\varphi}\xi''_{\alpha y}\nonumber\\
\xi_{\alpha y}&=&S_{\varphi}(C_{\gamma}\xi''_{\alpha x}-S_{\gamma}\xi''_{\alpha z})
+C_{\varphi}\xi''_{\alpha y}\nonumber\\
\xi_{\alpha z}&=&S_{\gamma}\xi''_{\alpha x}+C_{\gamma}\xi''_{\alpha z} \ .
\end{eqnarray}
Substitution of this transformation in (\ref{AFs}) gives 
the quadratic Lagrangian (\ref{s3})  as well as the following
cubic term (again we omit the double prime in $\xi$)
\begin{eqnarray}
\label{L33}
{\cal L}_3&=& 
\rho_s\left\{\left(\frac{1}{2}-b^2\right)n_z^2{\bm Q}
-\frac{Q}{2\sqrt{2}}\sum_{\alpha}\xi_{\alpha \perp}^2
{\bm e}_{\alpha} \right.\nonumber\\
&+&
\left. \frac{b^2}{2}\sum_{\alpha}({\bm \nabla}\xi_{\alpha y})n_z
\right\}{\bm \nabla}\varphi
\end{eqnarray}
\begin{figure}[h]
\includegraphics[width=0.2\textwidth,clip]{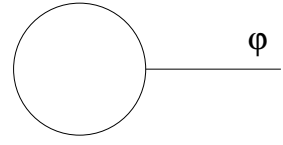}
\caption{The tadpole diagram giving the quantum correction
to the spin spiral wave vector.
}
\label{tpole}
\end{figure}
By calculating the tadpole with $\varphi$ as an external leg, 
see Fig.~\ref{tpole}, we find
\begin{eqnarray}
\label{tp}
&&{\cal L}_3 \to \rho_s ({\bm Q}\cdot{\bm \nabla})\varphi\\
&&\times\left\{\left(\frac{1}{2}-b^2\right)\langle n_z^2\rangle
-\frac{1}{2}\langle\xi_{\alpha \perp}^2\rangle
-\frac{b^2}{2Q^2}\langle \sum_{\alpha}\xi_{\alpha y}({\bm Q}
\cdot{\bm \nabla})n_z\rangle
\right\}\nonumber
\end{eqnarray}
Value of the wave vector of the spin spiral follows from the
condition that there are no linear in ${\bm \nabla}\varphi$
terms in the effective action.
The corresponding term in (\ref{AFs}) in the leading approximation is 
$-\rho_s({\bm Q}\cdot{\bm \nabla}\varphi)+\frac{g}{\sqrt{2}}
\sum_{\alpha}({\bm e}_{\alpha}\cdot{\bm \nabla}\varphi)$.
The condition of the disappearance of this term gives
the value of $Q$ presented in Eq. (\ref{Qn}).
The account of the tadpole (\ref{tp}) gives the  quantum correction
to the spiral vector,
\begin{eqnarray}
\frac{\delta Q}{Q}&=&\left(\frac{1}{2}-b^2\right)\langle n_z^2\rangle
-\frac{1}{2}\langle\xi_{\alpha \perp}^2\rangle\nonumber\\
&-&\frac{b^2}{2Q^2}\langle \sum_{\alpha}\xi_{\alpha y}({\bm Q}\cdot{\bm \nabla})
n_z\rangle \ .
\end{eqnarray}
The last term here is very similar to (\ref{xinr}),
\begin{eqnarray}
\label{xinr1}
&&\frac{1}{2Q^2}\langle \sum_{\alpha}
\xi_{\alpha y}({\bm Q}\cdot{\bm \nabla})
n_z\rangle  = \frac{cQ}{2\rho_s}{\cal K}'\\
&&{\cal K}'=\int\sum_{l=1,2}\frac{Z_{l,{\bm p}}}{2\Omega_{l,{\bm p}}}
\frac{2a^2[1+b^2p^2][1+b^2]p_1^2}
{[\Omega_{l,{\bm p}}^2-O_2]}\frac{d^2p}{(2\pi)^2} \ . \nonumber
\end{eqnarray}
It is also ultraviolet divergent. We discuss the divergence in the
next subsection. 

\subsection{Discussion and conclusion of the
``antiferromagnetic spiral'' section}
Both ${\cal K}$ in Eq.(\ref{xinr}) and ${\cal K}'$ in Eq.(\ref{xinr1})
are ultraviolet divergent. There is no a formal
way to renormalize out these ultraviolet divergences.
The effective bosonic field theory defined by the action (\ref{AFs}) is
unrenormalizable. 
Practically this means that  quantum corrections depend on the
ultraviolet cutoff and different physical quantities depend on this
cutoff in different ways. So it is impossible to reduce renormalization
of all physical quantities to the renormalization of a finite set of 
parameters.

Thus, the simplest conclusion is that within the
effective low energy bosonic theory (\ref{AFs}) one can calculate the spin-wave
spectra. The results are  presented in 
Figs.~\ref{IntFI},\ref{dispA05},\ref{dispA03}.
On the other hand it is impossible to calculate  quantum corrections.
However, we can somewhat improve the situation if recall that the effective
action (\ref{AFs}) originates from the effective action (\ref{eq:LL})
that is equivalent to the t-J model at low doping.
Therefore, the ultraviolet cutoff for fluctuations of the
$n$-field (\ref{nred1}) is about the inverse lattice spacing,
$\Lambda_n\sim 1/a$.  This cutoff does not appear in our
results explicitly as soon as we use the renormalized spin
stiffness and spin-wave velocity of the Heisenberg model~\cite{SZ}
$\rho_s=0.18J$, $c=1.65J$ (we set the lattice spacing equal to
unity, $a=1$).
The ultraviolet cutoff for fluctuations of the $\xi$-field
(\ref{xired1}), (\ref{xinr}), (\ref{xinr1})
is of the order of the Fermi momentum of holes, $\Lambda_{\xi}\sim
p_F \sim \sqrt{\pi x}$, where $x \ll 1$ is the doping level.
At $q > p_F$ description of fermions in terms of a structureless
bosonic field $\xi$ does not make sense.
The wave vector of the spin spiral scales linearly with 
doping~\cite{Milstein08}, $Q \propto x$.
Therefore, we conclude from Eqs. (\ref{xinr}) and (\ref{xinr1})
that the quantum fluctuation corrections to the 
spiral wave vector and to the pseudoscalar order parameter
are 
\begin{eqnarray}
\label{corr1}
\frac{\delta Q}{Q}\propto b^2\sqrt{x}\ , \ \ \ \ 
\frac{\delta {\cal P}}{{\cal P}}\propto +\sqrt{x} \ .
\end{eqnarray}
The $\sqrt{x}$ dependence originate from the integrals
${\cal K}$ and ${\cal K}'$ that are regularized by imposing
the ultraviolet cutoff  $\Lambda_{\xi}\sim \sqrt{ x}$.
There are two important points about Eqs.(\ref{corr1}).
{\it (i)} The quantum correction to the spiral wave vector is proportional
to $b^2$. According to experimental spectra~\cite{Tranq,hinkov04,hinkov07a} 
$b \lesssim 0.25$ and hence $b^2 \lesssim 0.06$.
So, we predict a very small quantum correction to the vector
of the spin spiral.
{\it (ii)} The sign of the $\delta {\cal P}$ is positive since ${\cal K}$ in
(\ref{xinr}) is negative. Hence quantum fluctuations enhance the
pseudoscalar order parameter. This trend is opposite to the trend
in $\langle {\vec n}\rangle$. According to Eq. (\ref{nred1})
quantum fluctuations reduce value of $|\langle {\vec n}\rangle|$
and at a sufficiently large $Q$ the static component of spin
vanishes, $|\langle {\vec n}\rangle|=0$.
Disappearance of static spin does not mean that the spin spiral
disappears, the spin spiral just becomes dynamic.
Hence, $|\langle {\vec n}\rangle|=0$ indicates
 the Quantum Critical Point (QCP) for
transition from the static spin spiral to the 
dynamic one. The spin spiral itself is characterised by the
pseudoscalar chiral order parameter ${\cal P}$  independently whether 
this is static
or dynamic case. This order parameter goes smoothly through the QCP.
So, the spin spiral survives in the magnetically disordered phase.

There are two types of spin wave excitations,
the in-plane magnons and the out-of-plane magnons.
The spectrum of the in-plane magnons is given by Eq.(\ref{i1}) and 
Fig.~\ref{IntFI}, and that of the out-of-plane magnons is given 
by Eq.(\ref{oi1}) and Figs.~\ref{dispA05},\ref{dispA03}.
It follows from Eq.(\ref{i1}) that neutron scattering
with excitation of in-plane magnons is necessarily accompanied by  
diffraction on the static spiral. 
Therefore, one should expect that the intensity of the
inelastic neutron scattering with excitation of in-plane magnons
is sharply decreasing at approaching the QCP. The intensity is zero in
the magnetically disordered phase.
On the other hand the intensity of neutron scattering with excitation
of out-of-plane magnons  
(Figs.~\ref{dispA05},\ref{dispA03} with the hourglass dispersion) 
is not  related to a diffraction
and hence it is not very sensitive to the QCP.
The two branches of the hourglass
dispersion originate from two different degrees of freedom, spin and 
pseudospin.
The lower branch of the hourglass has strongly anisotropic
momentum dependence with zeros at the Goldstone points 
${\bm q}=\pm{\bm Q}$, while the upper branch is practically isotropic.
This behavior agrees with the typical  dispersion observed in 
cuprates.~\cite{Tranq,hinkov04,hinkov07a}

\section{Spin spiral close to ferromagnet.
The field theory inspired by ${\text {FeSrO}}_3$}
The present section is inspired by neutron scattering data~\cite{Ulrich}
from FeSrO$_3$ and FeCaO$_3$ that indicate the spin spiral close to a collinear 
ferromagnet.
The wave length of the spin spiral in FeSrO$_3$ is about 10 lattice spacing. 
The compound is an insulator, therefore there is only a spin field
in the game, there is no a pseudospin, and there is no other relevant
dynamic variable. To understand the kinematic structure of the effective
action we refer to the minimal Heisenberg model suggested in Ref.~\cite{Giniyat}
In the present paper we consider generic scenarios, therefore
we address for simplicity the 2D case. The 3D generalization is straightforward.
\subsection{Effective action}
\begin{figure}[ht]
\includegraphics[width=0.3\textwidth,clip]{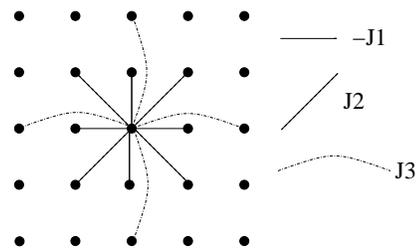}
\caption{2D Heisenberg $J_1-J_2-J_3$ model on square lattice.
There is spin ${\vec S}$ at each site.}
\label{s1}
\end{figure}

The Hamiltonian of the minimal Heisenberg model is
\begin{equation}
\label{H1}
H=-J_1\sum_{ij}{\vec S}_i\cdot{\vec S}_j
+J_2\sum_{i'j'}{\vec S}_{i'}\cdot{\vec S}_{j'}
+J_3\sum_{i''j''}{\vec S}_{i''}\cdot{\vec S}_{j''} \ .
\end{equation}
determined on the square lattice shown in Fig.~\ref{s1}. 
The $J_1$ link is ferromagnetic and $J_2,J_3$ links are antiferromagnetic.
For our purpose  it is sufficient to consider a classical spin, 
$S \gg 1$.
Let us impose the spiral configuration
\begin{equation}
\label{Ss}
{\vec S}=S(\cos({\bm Q}\cdot{\bm r}),\sin({\bm Q}\cdot{\bm r}),0) \ .
\end{equation}
The classical energy per unit cell corresponding to the Hamiltonian (\ref{H1})
reads
\begin{eqnarray}
\label{cE}
E&=&\frac{1}{a^2}S^2\left\{-J_1[\cos(Q_xa) +\cos(Q_ya)]\right.\nonumber\\
&+&2J_2\cos(Q_xa)\cos(Q_ya)\nonumber\\
&+&\left.J_3[\cos(2Q_xa) +\cos(2Q_ya)]\right\} \ ,
\end{eqnarray}
where $a$ is the lattice spacing.
Notice that in the present section we denote by indexes x and y
directions in the coordinate space.
Assuming that $Qa \ll 1$ and expanding the energy in powers of $Q$ we find
\begin{eqnarray}
\label{cE1}
E&=& -\frac{S^2}{2}Q^2(2J_2+4J_3-J_1)\\
&+&\frac{S^2J_2a^2}{2}Q_x^2Q_y^2
+S^2\frac{(16J_3+2J_2-J_1)a^2}{24}(Q_x^4+Q_y^4) \ .\nonumber
\end{eqnarray}
To have a small nonzero $Q$ we need a small {\it negative } spin stiffness
term. Thus, the effective energy per unit cell in terms of the 
ferromagnetic n-field, ${\vec n}^2=1$, reads
\begin{eqnarray}
\label{cEs}
E({\vec n})&=& -\frac{1}{2}\rho_s(\nabla{\vec n})^2\\
&+&b_1(\partial_x\partial_y{\vec n})^2
+\frac{1}{2}b_2[(\partial_x^2{\vec n})^2+(\partial_y^2{\vec n})^2] \ .\nonumber
\end{eqnarray}
The parameters are
\begin{eqnarray}
\label{J-s}
\rho_s&=&S^2(2J_2+4J_3-J_1)\nonumber\\
b_1&=&\frac{1}{2}S^2J_2a^2\nonumber\\
b_2&=&\frac{1}{12}S^2(16J_3+2J_2-J_1)a^2 \ .
\end{eqnarray}
It is clear that  the energy (\ref{cEs}) is
the most general expansion up to the fourth order derivatives.
This is true for the minimal Heisenberg model (\ref{H1}), and
this is also true for any extension of the Heisenberg model
that does not include mobile fermions. 
Depending on relation between the parameters $b_1$ and $b_2$ the spiral 
is directed along a diagonal or along an axis of the square lattice.
One can also tune up $b_1$ and $b_2$
in such a way that there is a directional degeneracy.
The ground state energy gain due to the spin spiral is proportional 
to $Q^4$.
All in all this is {\it very much}  different from the chiral scenario that
we have considered for the ``antiferromagnetic spiral''.
The effective action in the present case is
\begin{eqnarray} 
\label{Fsnn}
{\cal L}_{Fs}&=&i\frac{S}{a^2}\left(\eta^{\dag}{\dot \eta}
- {\dot \eta}^{\dag}\eta\right)-E({\vec n})\ .
\nonumber\\
{\vec n}&=&\eta^{\dag}{\vec \sigma}\eta \ . 
\end{eqnarray}
This is still a gradient expansion, but the  theory is 
qualitatively different from (\ref{AFs}).
The AF chiral theory (\ref{AFs}) originates from the t-J model
and therefore the wave vector of the spin spiral ${\bm Q}$
is naturally small, it is proportional to the small doping.
 In the ``ferromagnetic spiral'' described by the action (\ref{Fsnn})
the wave vector ${\bm Q}$ is small only due to accidental cancellation
of different contributions to the spin stiffness. The cancellation
results in a very small negative stiffness.

To be specific we consider the case $b_1 < b_2$. In this case ${\bm Q}$ is
directed along one of the diagonals, for example
\begin{equation}
\label{Qd}
{\bm Q}=Q\left(\frac{1}{\sqrt{2}},\frac{1}{\sqrt{2}}\right) \ .
\end{equation}
Minimization of the energy (\ref{cEs}) gives
\begin{equation}
\label{QA}
Q = \sqrt{\frac{\rho_s}{b_1+b_2}} \ .
\end{equation}

\subsection{Spin wave excitation, linear approximation}
Equation of motion generated by (\ref{Fsnn}) reads
\begin{equation}
2i\frac{S}{a^2}{\dot{\eta}}=\frac{\delta E}{\delta {\vec n}}
\frac{\delta {\vec n}}{\delta{\eta^{\dag}}}\ .
\end{equation}
This results in the following equation for ${\vec n}$
\begin{eqnarray}
\label{eqm}
{\dot n}_{\alpha}=\frac{a^2}{S}\epsilon_{\alpha\beta\gamma}
\left[{\hat D}n_{\beta}\right]n_{\gamma} \ .
\end{eqnarray}
The indexes $\alpha$, $\beta$, $\gamma$ denote components of
vector in the spin space.
 The differential operator in the square brackets acting on a plane
wave gives a polynomial,
\begin{eqnarray}
\label{polin}
{\hat D}&=&\frac{a^2}{S}\left[
\rho_s\Delta +2b_1\partial_x^2\partial_y^2
+b_2\left(\partial_x^4+\partial_y^4\right)\right]\nonumber\\
&&{\hat D}e^{i{\bm q}\cdot{\bm r}}=D_{\bm q}e^{i{\bm q}\cdot{\bm r}}\nonumber\\
D_{\bm q}&=&\frac{a^2}{S}\left[
-\rho_sq^2+2b_1q_x^2q_y^2
+b_2\left(q_x^4+q_y^4\right)\right]\ .
\end{eqnarray}
Looking for solution in the form
\begin{equation}
\label{QQQ}
{\vec n}\approx (\cos({\bm Q}\cdot{\bm r}+\varphi_{\bm q}),
\sin({\bm Q}\cdot{\bm r}+\varphi_{\bm q}), h_{\bm q})\ ,
\end{equation}
and substituting this in  Eq.(\ref{eqm})
we obtain the following equations
\begin{eqnarray}
\label{bq}
{\dot \varphi}_{\bm q}&=&-\left[D_{\bm Q}-D_{\bm q}\right]h_{\bm q}\nonumber\\
{\dot h}_{\bm q}&=&\left[D_{\bm Q}-\frac{1}{2}D_{\bm {q+Q}}
-\frac{1}{2}D_{\bm {q-Q}}\right]\varphi_{\bm q}\ .
\end{eqnarray}
The fixed-frequency solution of these equations, 
$\varphi_{\bm q},h_{\bm q}\propto exp(-i\omega_{\bm q}t)$,
results in the following frequency
\begin{eqnarray}
\label{fD}
\omega_{\bm q}=\sqrt{\left[\frac{1}{2}D_{\bm {q+Q}}
+\frac{1}{2}D_{\bm {q-Q}}-D_{\bm Q}\right]
\left[D_{\bm q}-D_{\bm Q}\right]}\ .
\end{eqnarray}
It is convenient to rewrite the frequency in the following notations
\begin{eqnarray}
\label{oqab}
\omega_{\bm q}&=&\Omega_0f_{\bm p}\\
f_{\bm p}&=&\sqrt{X_{\bm p}Y_{\bm p}}\ ,\nonumber
\end{eqnarray}
where $p$ is the dimensionless momentum defined in (\ref{dim}).
\begin{eqnarray}
\label{AB}
\Omega_0&=&(b_1+b_2)\frac{Q^4a^2}{S}\ ,\nonumber\\
X_{\bm p}&=&2(p_{1}^2+\alpha p_{2}^2)+\frac{1}{2}p^4+2\alpha p_{1}^2p_{2}^2
\ ,\nonumber\\
Y_{\bm p}&=&\frac{1}{2}(1-p^2)^2+2\alpha p_{1}^2p_{2}^2\ ,\nonumber\\
\alpha&=&\frac{b_2-b_1}{b_2+b_1} \ .
\end{eqnarray}
We remind that $p_1$ is directed along ${\bm Q}$ and $p_2$
is perpendicular to ${\bm Q}$.
\begin{figure}[h]
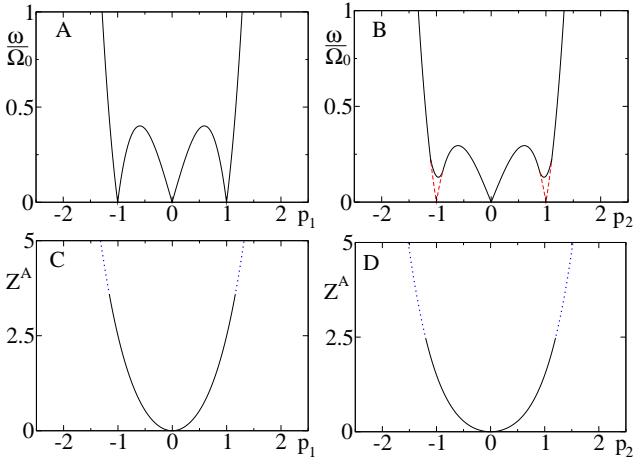

\includegraphics[width=0.23\textwidth,clip]{ferrodisp1.eps}
\includegraphics[width=0.23\textwidth,clip]{ferrodisp2.eps}
\includegraphics[width=0.23\textwidth,clip]{intA1.eps}
\includegraphics[width=0.23\textwidth,clip]{intA2.eps}
\caption{(Color online) A and B: the magnon dispersion 
in the direction  parallel to the spin spiral (A), ${\bm p}=(p_1,0)$,
and in the direction perpendicular to the spin spiral (B),
${\bm p}=(0,p_2)$.
The red dashed line in Fig. B shows schematically opening of the gap
driven by quantum fluctuations as it is discussed in
subsection D of the present section.
 The ``unshifted'' intensity factors  $Z^{(A)}_{\bm k}=\frac{\omega_{\bm k}}
{\Omega_0} A_{\bm k}^2$ 
are plotted in Figs. (C) and (D) for the same directions.
The intensity factors at values of ${\bm k}$ where
$\omega_{\bm k} > 0.5\Omega_0$ are shown by blue dotted lines.
All the plots are presented for the anisotropy parameter $\alpha=0.5$.
}
\label{ferrodisp}
\end{figure}
The dispersion (\ref{oqab}) is plotted
at $\alpha=0.5$ for directions parallel to the spin spiral, 
Fig.~\ref{ferrodisp}A,
and perpendicular to the spin spiral, Fig.~\ref{ferrodisp}B. 
There are four points to note.
{\it (i)} The dispersion scales as $\Omega_0 \propto Q^4$
that is a consequence of the fourth derivatives in the action.
{\it (ii)} The energy is zero at ${\bm p}=0$ and at ${\bm p}=(\pm 1,0)$.
    These zeroes are dictated by the Goldstone theorem.
{\it (iii)} The energy is also zero at ${\bm p}= (0,\pm 1)$.
These zeroes are not dictated by any exact theorem and we will see
that quantum fluctuations open a gap at these points.
{\it (iv)} The dispersion at  $p \gg 1$ is anisotropic,
$\omega_{\bm q}\approx \Omega_0\left[p^4/2+2\alpha p_1^2p_2^2\right]$.
This is because the theory is stabilized by forth spacial derivatives 
that ``know'' about the lattice structure.

Representing $\varphi$ and $h$ in terms of the spin wave
annihilation   and creation  operators ($c_{\bm p}$, $c_{\bm p}^{\dag}$),
\begin{eqnarray}
\label{bg12}
h&=&\sum_{\bm q}-iA_{\bm q}c_{\bm q}e^{-i\omega_{\bm q}t+i{\bm q}\cdot{\bm r}}+h.c.
\nonumber\\
\varphi&=&\sum_{\bm q}B_{\bm q}c_{\bm q}e^{-i\omega_{\bm q}t+i{\bm q}\cdot{\bm r}}+h.c. \ ,
\end{eqnarray}
and applying the standard quantization procedure to the action (\ref{Fsnn})
we find the coefficients $A_{\bm p}$ and $B_{\bm p}$,
\begin{eqnarray}
\label{AB1}
A_{\bm p}&=&\frac{\sqrt{X_{\bm p}}}{\sqrt{2S f_{\bm p}}} \ ,\nonumber\\
B_{\bm p}&=&\frac{\sqrt{Y_{\bm p}}}{\sqrt{2S f_{\bm p}}} \ .
\end{eqnarray}

\begin{figure}[h]
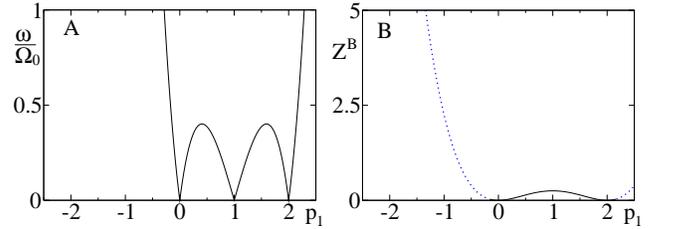

\includegraphics[width=0.23\textwidth,clip]{dispS.eps}
\includegraphics[width=0.23\textwidth,clip]{intB1.eps}
\caption{(Color online) 
A: the positively shifted magnon dispersion $\omega_{\bm {k-Q}}$
for the direction of ${\bm k}$ parallel to the spiral.
B: the intensity factor $Z^{(B)}_{\bm k}=\frac{\omega_{\bm {k-Q}}}{2\Omega_0} 
B_{\bm {k-Q}}^2$ 
for the positively shifted line.
The intensity factor at values of ${\bm k}$ where
$\omega_{\bm {k-Q}} > 0.5\Omega_0$ is shown by blue dotted line.
The plots are presented for the anisotropy parameter $\alpha=0.5$.
}
\label{ferrodisp1}
\end{figure}

Neutron scattering amplitude is given by the Fourier transform of (\ref{QQQ})
and hence the scattering probability with the
energy transfer $\omega$ and the momentum transfer ${\bm k}$
is proportional to
\begin{eqnarray}
\label{fI}
&&I(\omega,{\bm k})={\cal R}_{\perp} A_{\bm k}^2\delta(\omega-\omega_{\bm k})\\
&& \ \ \ +\frac{1}{2}{\cal R}_{||}\left\{
B_{\bm {k+Q}}^2\delta(\omega-\omega_{\bm {k+Q}})
+B_{\bm {k-Q}}^2\delta(\omega-\omega_{\bm {k-Q}})\right\} \ ,\nonumber
\end{eqnarray}
where $ {\cal R}_{\perp}$ and ${\cal R}_{||}$ are kinematic factors
that depend on the relative orientation of the neutron polarization
and the plane of the spin spiral.
These factors depend on  specific experimental details and
therefore we do not present these factors.
Thus, there is an ``unshifted'' contribution in the neutron scattering
probability with intensity proportional to $A_{\bm k}^2$ and two
``shifted'' contributions with intensities proportional to 
$\frac{1}{2}B_{\bm {k\pm Q}}^2$. As one expects, the intensity
is diverging at points where $\omega_{\bm k} \to 0$.
To characterise the intensity we plot the intensity factor
$Z^{(A)}_{\bm k}=\frac{\omega_{\bm k}}{\Omega_0} A_{\bm k}^2$ in
Figs.~\ref{ferrodisp}C and Figs.~\ref{ferrodisp}D
for directions of ${\bm k}$ parallel and  perpendicular to the spiral.
The ``positively shifted'' spectrum $\omega_{\bm {k-Q}}$ is plotted
in Fig.~\ref{ferrodisp1} for the direction of ${\bm k}$
parallel to the spiral.
In the same Fig.~\ref{ferrodisp1} we also plot the
 intensity factor $Z^B_{\bm k}=\frac{\omega_{\bm {k-Q}}}{2\Omega_0} B_{\bm {k-Q}}^2$
for the shifted line.
Comparing Figs.~\ref{ferrodisp}~C,D and Fig.~\ref{ferrodisp1}B
we conclude that the unshifted ``line'' has a significant
amplification compared to the shifted one.

\subsection{Quantum and thermal fluctuations. 
Dependence of the spiral wave vector on temperature}
Since we are interested in nonlinear corrections, Eq. (\ref{QQQ})
has to be written more accurately,
\begin{eqnarray}
\label{Qnbg}
{\vec n}&=&(C_{\varphi},S_{\varphi},0)\cos\gamma
+(0,0,1)\sin\gamma \nonumber\\
C_{\varphi}&=&\cos({\bm Q}\cdot{\bm r}+\varphi) \nonumber\\
S_{\varphi}&=&\sin({\bm Q}\cdot{\bm r}+\varphi) \nonumber\\
h&=&\sin\gamma \ .
\end{eqnarray}
Equations (\ref{Qd}) and (\ref{QA}) have been derived 
by minimization of the semiclassical energy.
Alternatively they can be obtained from the
condition that the linear in operators term in the action (energy),
\begin{eqnarray}
\label{lc}
E_1&=&-\rho_s\left(Q_x\partial_x\varphi+Q_y\partial_y\varphi\right)\nonumber\\
&+&2b_1Q_xQ_y\left(Q_y\partial_x\varphi+Q_x\partial_y\varphi\right)\nonumber\\
&+&2b_2\left(Q_x^3\partial_x\varphi+Q_y^3\partial_y\varphi\right)\ ,
\end{eqnarray}
is zero, $E_1=0$.
Using (\ref{Qnbg}) and expanding Eq. (\ref{cEs}) up to cubic terms 
in perturbations $h$ and $\varphi$ we find the following cubic terms
\begin{eqnarray}
\label{c3}
({\bm \nabla}{\vec n})^2&\to&-2 h^2({\bm Q}\cdot{\bm {\nabla}})\varphi \\
(\partial_{\mu}\partial_{\nu}{\vec n})^2&\to&2
\left(Q_{\mu}\partial_{\nu}\varphi+Q_{\nu}\partial_{\mu}\varphi\right)
\left[(\partial_{\mu}\varphi)(\partial_{\nu}\varphi)\right.\nonumber\\
&-&\left.Q_{\mu}Q_{\nu}h^2
+2\partial_{\mu}\left(h\partial_{\nu}h\right)
\right] \ .\nonumber
\end{eqnarray}
There is no summation over $\mu$ and $\nu$ in (\ref{c3}), so one
can set $\mu=x$, $\nu=y$ or $\mu=x$, $\nu=x$, etc.
To calculate the tadpole, Fig.~\ref{tpole}, with interaction
(\ref{cEs}),(\ref{c3}) we note  that the following
expectation values are nonzero
\begin{eqnarray}
\label{expe}
{\cal M}&=&\langle (\partial_x\varphi) (\partial_x\varphi)\rangle=
\langle (\partial_y\varphi) (\partial_y\varphi)\rangle\nonumber\\
&=&a^2\int\frac{d^2q}{(2\pi)^2}
B_{\bm p}^2(q_1^2+q_2^2)\left(n_{\bm q}+\frac{1}{2}\right)
\nonumber\\
{\cal N}&=&\langle (\partial_x\varphi) (\partial_y\varphi)\rangle\nonumber\\
&=&a^2\int\frac{d^2q}{(2\pi)^2}B_{\bm p}^2(q_1^2-q_2^2)
\left(n_{\bm q}+\frac{1}{2}\right)\ ,\nonumber\\
\langle h^2\rangle&=&
\int\frac{d^2q}{(2\pi)^2}2A_{\bm p}^2\left(n_{\bm q}+\frac{1}{2}\right) \ .
\end{eqnarray}
Here 
\begin{equation}
n_{\bm q}=\frac{1}{e^{\omega_{\bm q}/T}-1}
\end{equation}
is the usual Bose distribution. 
We remind also that the dimensionless $p$ is related to $q$
by Eq.(\ref{dim}), and note that 
$\langle \partial_{\mu}(n_z\partial_{\nu}n_z)\rangle=0$.

When calculating the tadpole the Hartree-Fock decoupling in 
Eqs.(\ref{cEs}) and (\ref{c3}) gives two kind of terms.
Terms of the first kind proportional to $\langle n_z^2\rangle$ lead to 
the following renormalization
\begin{eqnarray}
\label{r1}
\rho_s&\to&\rho_s(1-\langle n_z^2\rangle) \nonumber\\
b_1&\to&b_1(1-\langle n_z^2\rangle) \nonumber\\
b_2&\to&b_2(1-\langle n_z^2\rangle) \ .
\end{eqnarray}
Obviously, this renormalization does not change $Q$ given by Eq.(\ref{QA}).
Terms of the second kind are proportional to ${\cal M}$ and ${\cal N}$ defined in
(\ref{expe}).
They generate the following fluctuation correction (tadpole)
to the effective energy
\begin{eqnarray}
\label{mn}
E_3&=&2{\cal M}(b_1+3b_2)(Q_x\partial_x\varphi+Q_y\partial_y\varphi)\nonumber\\
&+&4{\cal N}b_1(Q_x\partial_y\varphi+Q_y\partial_x\varphi) \ .
\end{eqnarray}
The renormalized spiral wave vector $Q$ has to be found from the condition $E_1+E_3=0$,
see Eqs. (\ref{lc}),(\ref{mn}).
This gives the following correction to $Q$
\begin{eqnarray}
\label{dQ}
&&\hspace{-10pt}\frac{\delta Q}{Q}=-\frac{1}{Q^2}\sum_{\bm q}B_{\bm q}^2\left[3q_1^2+(1+2\alpha)q_2^2\right]
\left(n_{\bm q}+\frac{1}{2}\right)\\
&&=
-\frac{Q^2a^2}{2S}\int \frac{d^2p}{(2\pi)^2}\sqrt{\frac{Y_{\bm p}}{X_{\bm p}}}
\left[3p_1^2+(1+2\alpha)p_2^2\right]
\left(n_{\bm p}+\frac{1}{2}\right) \nonumber
\end{eqnarray}
We remind again that the momentum $p$ in this equation is dimensionless,
 $p = q/Q$.
The quantum fluctuation part of $\delta Q$,
that comes from $1/2$ in $\left(n_{\bm p}+1/2\right)$,
is ultraviolet divergent as $p^4$  (in 3D it would be $p^5$).
Therefore it must be renormalized to zero. The temperature dependence 
of $Q$ is given by the
$n_{\bm p}$-term that is left after the renormalization.

To understand better the quantum renormalization procedure we rewrite
(\ref{lc}) and (\ref{mn}) having in mind that $Q_x=Q_y=Q/\sqrt{2}$,
\begin{eqnarray}
\label{E13}
E_1&=&\left[-\rho_sQ+(a_1+a_2)Q^3\right]\partial_1\varphi\nonumber\\
E_3&=&\left[2{\cal M}(a_1+3a_2)Q+4{\cal N}a_1Q\right]\partial_1\varphi \ .
\end{eqnarray}
Using (\ref{expe}) one can easily calculate ${\cal M,N}$ at T=0,
\begin{eqnarray}
\label{MNl}
{\cal M}&=&\frac{a^2}{2S}\left[\frac{\Lambda^4}{16\pi}
+f_2\Lambda^2Q^2
+f_4 Q^4\ln\left(\frac{\Lambda}{Q}\right)
+f_6Q^4\right]\nonumber\\
{\cal N}&=&\frac{a^2}{2S}\left[
g_2\Lambda^2Q^2
+g_4Q^4\ln\left(\frac{\Lambda}{Q}\right)
+g_6Q^4\right] \ ,
\end{eqnarray}
where $\Lambda \sim 1/a$ is the ultraviolet cutoff and
$f_2,f_4,f_6,g_2,g_4,g_6$ are some functions of the anisotropy 
$\alpha$ that can be easily calculated.
Comparing (\ref{MNl}) with (\ref{E13}) we see that the $\Lambda^4$
term renormalizes $\rho_s$ and the $\Lambda^2Q^2$ terms renormalize
$(a_1+a_2)$.
The $Q^4\ln\left(\frac{\Lambda}{Q}\right)$ and $Q^4$ terms
give contributions to the energy density of the order of 
$\sim \frac{J}{S}(Qa)^4Q^2$.
So, the $Q^4\ln\left(\frac{\Lambda}{Q}\right)$ and $Q^4$
terms in (\ref{MNl}) 
 have to be neglected since  in (\ref{cEs}) we keep only the terms
$\sim JQ^2$ and $\sim J(Qa)^2Q^2$ neglecting higher derivatives.

\begin{figure}[h!]
\includegraphics[width=0.22\textwidth,clip]{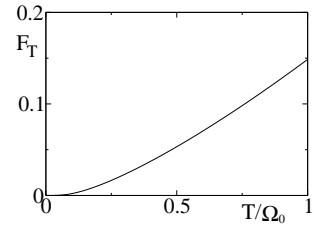}
\caption{Temperature dependence of the integral $F_T$ in
Eq.(\ref{dQT}) for $Q(T)$. The value of $\alpha$ is 0.5}
\label{QT}
\end{figure}
Temperature dependence of the spiral wave vector is determined
by the ``$n_{\bm p}$-part'' of Eq. (\ref{dQ}),
\begin{eqnarray}
\label{dQT}
\frac{\delta Q}{Q}&=&-\frac{Q^2a^2}{2S} \ F_T\\
F_T&=&\int \frac{d^2p}{(2\pi)^2}\sqrt{\frac{Y_{\bm p}}{X_{\bm p}}}
\left[3p_1^2+(1+2\alpha)p_2^2\right] n_{\bm p} \ .\nonumber
\end{eqnarray}
The dimensionless integral $F_T$ is plotted in Fig.\ref{QT}
versus $T/\Omega_0$ for $ \alpha=0.5$.
Temperature reduces the value of $Q$ and this is a sizable effect.

\subsection{Quantum correction to the spin-wave dispersion.
Opening of the gap at ${\bm q}=(0,\pm Q)$}
\begin{figure}[h]
\includegraphics[width=0.15\textwidth,clip]{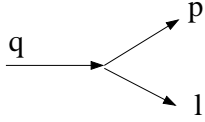}
\caption{The tree leg vertex}
\label{3leg1}
\end{figure}

\begin{figure}[h!]
\includegraphics[width=0.45\textwidth,clip]{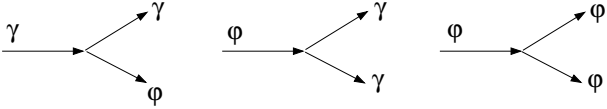}
\caption{Three different contributions to the 3-leg vertex}
\label{3leg2}
\end{figure}

There is the tree leg vertex in the theory shown in Fig.\ref{3leg1} 
The vertex is  determined by Eqs.(\ref{cEs}) and (\ref{c3}).
There are three contributions to the vertex.
The first contribution comes from the $\gamma \to \gamma \varphi$ terms
shown in Fig.\ref{3leg2} left. 
The second contribution comes from the
$\varphi \to \gamma \gamma$ terms, Fig.\ref{3leg2} middle,
 and the third contribution comes from the 
$\varphi \to \varphi \varphi$ terms, Fig.\ref{3leg2} right.
A straightforward calculation gives the following answers for
these contributions
\begin{eqnarray}
\label{vertex3leg}
\Gamma&=&\Gamma_1+\Gamma_2+\Gamma_3 \\
\Gamma_1&=&2iQ(a_1+a_2)\times \left\{
p_1\left[p_1^2+p_2^2(1+2\alpha)\right]A_{\bm q}B_{\bm p}A_{\bm l}
\right.\nonumber\\
&+&\left.
l_1\left[l_1^2+l_2^2(1+2\alpha)\right]A_{\bm q}B_{\bm l}A_{\bm p}\right\}
\nonumber\\
\Gamma_2&=&2iQ(a_1+a_2)\times q_1\left[q_1^2+q_2^2(1+2\alpha)\right]
B_{\bm q}A_{\bm p}A_{\bm l}\nonumber\\
\Gamma_3&=&-2iQ(a_1+a_2)\times \left\{3q_1p_1l_1\right.\nonumber\\
&+&\left.(1+2\alpha)
\left[q_1p_2l_2+p_1q_2l_2+l_1p_2q_2\right]\right\}B_{\bm q}B_{\bm p}B_{\bm l} \ .
\nonumber
\end{eqnarray}
Importantly, for an {\it on shell} scattering process, 
${\bm q}={\bm p}+{\bm l}$, $\omega_{\bm q}=\omega_{\bm p}+\omega_{\bm l}$, 
the vertex satisfies the Adler's relation, it vanishes 
if one of the momenta equals to $0$ or ${\bm Q}$.

\begin{figure}[h]
\includegraphics[width=0.4\textwidth,clip]{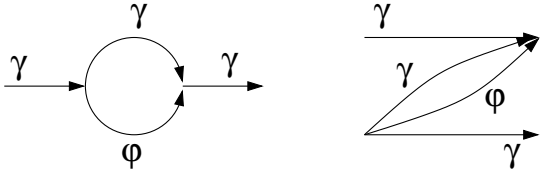}
\caption{The $\gamma-\gamma$ polarization operator
}
\label{loopgg}
\end{figure}

To understand the structure  of the quantum correction to
the spin-wave dispersion we calculate the $\gamma-\gamma$
polarization operator shown in Fig.\ref{loopgg}
\begin{eqnarray}
\label{pgg}
P^{\gamma\gamma}_{\bm q}(\omega)=\sum_{\bm k}
\frac{2\Gamma_1^*\Gamma_1(\omega_{{\bm q}/2-{\bm k}}+\omega_{{\bm q}/2+{\bm k}})}
{\omega^2-(\omega_{{\bm q}/2-{\bm k}}+\omega_{{\bm q}/2+{\bm k}})^2}
\end{eqnarray}
The polarization operator can be represented in the following form
\begin{eqnarray}
\label{pgg1}
P^{\gamma\gamma}_{\bm q}(\omega)= -A_{\bm q}^2\frac{Q^2(a_1+a_2)}{S}a^2
\times integral \ .
\end{eqnarray}
The {\it integral} is quadratically ultraviolet diverging.
Similarly to (\ref{MNl}) only ultraviolet diverging 
contributions are important  with the accepted accuracy.
Calculation of diverging terms in the {\it integral}
is very simple. For example the result at $\alpha=0.5$  is
\begin{eqnarray}
\label{int2}
&&integral \to (5.2q_1^2+1.5q_2^2)\frac{\Lambda^2}{4\pi}
+\frac{\ln(\Lambda/Q)}{2\pi}\\
&&\times(8.7Q^2q_1^2+5.0Q^2q_2^2
-4.7q_1^4-1.4q_2^4-10q_1^2q_2^2)
\nonumber
\end{eqnarray}
The polarization operator determines the 
$\gamma_{\bm q}^{\dag}\gamma_{\bm q}$ term in  the effective action.
The coefficient $A_{\bm q}^2$ in (\ref{pgg1}) gives normalization of
the $\gamma$-field and powers of momentum in (\ref{int2}) 
should be replaced by corresponding derivatives
of this field. All in all Eq. (\ref{pgg1}) together
with Eq. (\ref{int2}) result in the following effective energy
\begin{eqnarray}
\label{degg}
\delta E_{\gamma\gamma}&=&\frac{1}{S}Q^2(a_1+a_2)\\
&\times&\left\{
-\left[5.2\frac{(\Lambda a)^2}{4\pi}+
8.7(Qa)^2\frac{\ln(\Lambda/Q)}{2\pi}\right](\partial_1\gamma)^2\right.\nonumber\\
&-&\left.
\left[1.5\frac{(\Lambda a)^2}{4\pi}+
5.0(Qa)^2\frac{\ln(\Lambda/Q)}{2\pi}\right](\partial_2\gamma)^2
\right.\nonumber\\
&+&\frac{\ln(\Lambda/Q)}{2\pi}a^2
[4.7(\partial_1^2\gamma)^2+1.4(\partial_2^2\gamma)^2\nonumber\\
&+&\left.10(\partial_1\partial_2\gamma)^2]\right\}\nonumber
\end{eqnarray}
All momenta in the physically interesting region
are of the order of $Q$. Hence, the typical
value of the bare energy density (\ref{cEs}) is 
$E_{bare}\sim \frac{S\Omega_0}{a^2}$.
The logarithmic terms in (\ref{degg}) are of the order of
$\sim \frac{\Omega_0}{a^2}(Qa)^2$, so they must be neglected
within the accepted long-wave-length approximation, $Qa \ll 1$
Only the power divergent terms in (\ref{degg}) are significant
within the long-wave-length approximation, hence
\begin{eqnarray}
\label{degg1}
\delta E_{\gamma\gamma}\to-\frac{Q^2(a_1+a_2)}{S}\frac{(\Lambda a)^2}{4\pi}
\left[5.2(\partial_1\gamma)^2+1.5(\partial_2\gamma)^2\right] \ \
\end{eqnarray}
The structure of $\delta E_{\varphi\varphi}$ is the same.
The most important issue is that the polarization
operator correction (\ref{degg1}) contains different coefficients
in front of the derivatives in different directions.
To satisfy the Goldstone theorem we must introduce a contra-term 
$-\frac{1}{2}\delta \rho_s({\bm \nabla}{ \vec n})^2$
to renormalize out the longitudinal derivative, $(\partial_1\gamma)^2$.
Because of the mentioned different coefficients the term
$(\partial_2\gamma)^2$ is not zero after the renormalization.
Therefore, we necessarily open a gap in the dispersion at the points
${\bm q}=(0,\pm Q)$. The gap comes from fluctuations at the ultraviolet 
cutoff, therefore the effective field theory does not allow to calculate
the gap. An estimate for the gap follows from (\ref{degg1}),
\begin{equation}
\label{dsw}
\Delta \sim \frac{\Omega_0}{\pi S} \ .
\end{equation}
Here we take into account that $\Lambda \sim 1/a$.
The gap opening is shown  schematically in Fig.~\ref{ferrodisp}B.
Due to quantum fluctuations the gapless red dashed part of the dispersion
disappears and the dispersion becomes gaped.
It is worth noting that the mechanism of the gap opening
due to quantum fluctuations is similar to that in the spin-stripe phase of
the $J_1-J_2$ model.~\cite{Singh03,Uhrig09}

\subsection{Discussion and conclusion of the
``ferromagnetic spiral'' section}

We have formulated the effective field theory describing a spin spiral
close to a collinear ferromagnetic state, i.e. the wave vector of the
spiral is small compared to inverse lattice spacing, $Q \ll \pi/a$.
The problem is inspired by recent data~\cite{Ulrich} on FeSrO$_3$ and
FeCaO$_3$. We assume that there are only localized spins, no mobile
spin carriers. This assumption results in the effective action
(\ref{Fsnn}) that contains {\it forth} spacial derivatives.
Using the effective action we calculate the spin-wave excitation spectra,
see Figs.\ref{ferrodisp}~A,B.
The spectra have three zero modes at ${\bm q}=0,\ \ \pm{\bm Q}$
dictated by the Goldstone theorem.
In the semiclassical approximation there are also two additional
zero modes that are not dictated by any exact theorem.
We demonstrate that quantum fluctuations open gaps at these points.

We have also calculated the temperature dependence of the spiral wave vector,
 $Q=Q(T)$, see Eq.(\ref{dQT}) and Fig.~\ref{QT}.

The typical energy scale for the spin-wave excitation with momentum 
$q \sim Q$ is $\Omega_0 \propto Q^4$, see Eq.(\ref{AB}).
The dependence $\propto Q^4$ is a consequence of the forth order
derivatives in (\ref{Fsnn}).
Based on data~\cite{Ulrich} we estimate that   $\Omega_0 \sim 10$meV
in FeSrO$_3$ and FeCaO$_3$.
Therefore an external  magnetic field of the order of several
Tesla can substantially modify the spin spiral making it conical.

Last but not least. The intensity of
inelastic neutron scattering grows monotonously at large q, 
see Figs.\ref{ferrodisp}~B,C.
Within the accepted model we do not see a way to suppress intensity
outside of the region $q \sim Q$. We did calculate broadening
of magnons due to the decay 1 $\to$ 2, see Fig.~\ref{3leg1}, and
found that the broadening is very small. 
Introduction of mobile spins in addition to the localized ones
can change the situation. In this case a redistribution of the
quasiparticle weight between two brunches of the dispersion
similar to that for the AF spin spiral is possible. 
However, discussion of this effect is outside of the scope of the
present paper.

\section{Conclusion}
In the present paper we  derive and analyze effective quantum field theories
for helical spin magnets close to antiferromagnets and to
ferromagnets. We assume that the wave vector of the spin spiral
is  small compared to the wave vector of the lattice,
$Q \ll 2\pi/a$.
The noncollinear structure is caused by spin frustration.
The field theory describing the ``antiferromagnetic spin spiral''
is inspired by physics of cuprate superconductors.
In this case the wave vector of the spin spiral is naturally 
small because of the small doping $x$.
The field theory describing the ``ferromagnetic spin spiral''
is inspired by the recently studied FeSrO$_3$ and FeCaO$_3$ compounds.
In this case the wave vector of the spin spiral is small
due to accidental fine tuning of parameters.

The effective action describing the AF spin  spiral (\ref{AFs})
contains two bosonic fields, ${\vec n}$ corresponding to the staggered
spin, and ${\vec \xi}$ corresponding to pseudospin. The action is
quadratic in spacial derivatives, contains the second order time
derivative of ${\vec n}$ and the first order time derivative of the
spinor field describing ${\vec \xi}$ in the CP$^1$ representation.
The action is space and time covariant, the covariance is a consequence
of the gauge invariance of the t-J model which is the origin of
the effective action.
The  ``hourglass'' dispersion of magnons, see 
Figs. \ref{dispA05}, \ref{dispA03},
is a direct consequence of the action.
The two branches of the ``hourglass''
dispersion originate from two different degrees of freedom, spin and 
pseudospin.
The lower branch of the ``hourglass'' has strongly anisotropic
momentum dependence with two zero modes at the Goldstone points 
${\bm q}=\pm{\bm Q}$, while the upper branch is practically isotropic.
This behavior agrees with the typical  dispersion observed in 
cuprates.~\cite{Tranq,hinkov04,hinkov07a} 
Dependent on parameters of the action the AF spin spiral can be either static 
or dynamic.
There is a Quantum Critical Point that separates the static and the
dynamic regimes.
The spin spiral itself is characterised by the
pseudoscalar chiral order parameter independently whether 
this is static or dynamic case. The chiral order parameter changes sign
under time inversion.

The effective action of the ferromagnetic spin  spiral, Eqs. (\ref{Fsnn})
and (\ref{cEs}), contains only one bosonic field ${\vec n}$ describing spins.
The action contains the second and the fourth order spacial derivatives
as well as the first order time derivative of the
spinor field describing ${\vec n}$ in the CP$^1$ representation.
Dependent on parameters of the action the wave vector of the spin 
spiral ${\bm Q}$ can be directed either along a diagonal or along an axis
of a square (cubic) lattice. 
The effective action results in the spin-wave excitation spectra
shown in Figs.\ref{ferrodisp}~A,B.
The spectra have three zero modes at ${\bm q}=0,\  \pm{\bm Q}$
dictated by the Goldstone theorem.
In the semiclassical approximation there are also two additional
zero modes. Quantum fluctuations open gaps at these points.
The  wave vector of the spiral has a sizable temperature dependence that
has been calculated.

\acknowledgments
We are very grateful to C.~Ulrich and G.~Khaliullin for communicating their
results prior to publications.
Discussions with them as well as with D. Efremov were very important
for a significant part of this work.
A.~I.~M. gratefully acknowledges the School of Physics at the University of New
South Wales for warm hospitality and financial support during his visit.
This work was supported in part by the Australian Research Council.

\end{document}